\documentclass[11pt]{article}
\usepackage{amssymb}
\usepackage{graphics}
\usepackage{epsfig}
\usepackage{a4wide}
\usepackage{slashed}

\begin{document}

\newcommand{\ppqq}{$pp \to q_- q^{\prime}_- + X~$}
\newcommand{\qqqq}{$q q^{\prime} \to q_- q^{\prime}_-~$}
\newcommand{\uuqq}{$uu \to u_- + u_-~$}
\newcommand{\qqqqg}{$q q^{\prime} \to q_- + q^{\prime}_- + g~$}
\newcommand{\qgwwq}{$q(\bar q)g \to W_H^+ W_H^- + q(\bar q)~$}

\title{NLO QCD corrections to same-sign $T$-odd quark pair production
in the littlest Higgs model with $T$-parity at the LHC}
\author{ Xiong Shou-Jian, Ma Wen-Gan, Guo Lei, Chen Chong, and Zhang Ren-You\\
{\small  Department of Modern Physics, University of Science and Technology}  \\
{\small  of China (USTC), Hefei, Anhui 230026, P.R.China}}

\date{}
\maketitle \vskip 15mm
\begin{abstract}
We present the calculations for the same-sign $T$-odd mirror quark
pair production of the first two generations in the littlest Higgs model
with $T$-parity (LHT) at the $\sqrt{s}=14~{\rm TeV}$ LHC up to the QCD
next-to-leading order (NLO) including the subsequent decays of the $T$-odd
mirror quarks. The uncertainties from the factorization/renormalization
scale and parton distribution functions (PDFs) are discussed. Our
numerical results show that the PDF uncertainty of the NLO QCD corrected
cross section for the same-sign $T$-odd mirror quark pair production
of the first two generations is comparable with the scale uncertainty,
and the combined uncertainty at the QCD NLO is much smaller than that at
the LO with the factorization/renormalization scale $\mu$ in the range of $[\mu_0/4,~4\mu_0]$.
We also study the dependence of the total cross section on the LHT
parameters, and provide the transverse momentum, rapidity, invariant mass
and $H_T$ distributions of final products.
\end{abstract}


\vfill \eject \baselineskip=0.32in

\renewcommand{\theequation}{\arabic{section}.\arabic{equation}}
\renewcommand{\thesection}{\Roman{section}.}
\newcommand{\nb}{\nonumber}

\newcommand{\Dir}{\kern -6.4pt\Big{/}}
\newcommand{\Dirin}{\kern -10.4pt\Big{/}\kern 4.4pt}
\newcommand{\DDir}{\kern -7.6pt\Big{/}}
\newcommand{\DGir}{\kern -6.0pt\Big{/}}

\makeatletter      
\@addtoreset{equation}{section}
\makeatother       

\par
\section{Introduction}
\par
Although the standard model (SM) \cite{s1,s2} has been repeatedly
confirmed by high energy experiments, there are still a number of
theoretical problems unsolved, providing strong motivations to
search for physics beyond the SM. As one of them, the fine-tuning
problem of the Higgs boson \cite{fine-tuning} has become even more
attractive after the discovery of the $126~{\rm GeV}$ Higgs boson
by the ATLAS and CMS collaborations \cite{HiggsA,HiggsC}.
To regulate quadratically divergent contributions to the Higgs mass,
many extended models are proposed. Among them, the little Higgs
models \cite{Little Higgs models} deserve attention due to their
constructions as one kind of electroweak symmetry breaking (EWSB)
without fine-tuning problem. The Higgs boson in these models
manifests as a Nambu-Goldstone boson corresponding to a new global
symmetry, which guarantees the lightness of the Higgs boson through
its spontaneous breaking at a higher scale $f$.

\par
The most economical version of the little Higgs models is the
littlest Higgs model (LHM) \cite{the LH}. In addition to the SM
particle spectrum, a set of new heavy gauge bosons
$(W_H^{\pm},Z_H,A_H)$, a colored vector-like quark $(T)$ and a
scalar triplet $(\Phi)$ are introduced in the
LHM. Unfortunately, the original construction of the LHM conflicts
severely with precision electroweak constraints
\cite{EWconstraints}, forcing the value of $f$ to be as large as
several TeV. The fine-tuning between the cutoff scale $4 \pi f$
and the weak scale will be needed again for a too large $f$ value.
This problem can be solved
naturally when the LHM is endowed a discrete symmetry named
$T$-parity \cite{$T$-parity1, $T$-parity2, $T$-parity3}. In the
littlest Higgs model with $T$-parity (LHT), newly introduced
particles are odd under $T$-parity except $T_+$, while the SM
particles are $T$-even. $T$-parity conservation makes the $T$-odd
particles can only be produced in pairs, and the SM gauge
bosons can not mix with their $T$-partners. Hence the new particles
do not contribute to the electroweak precision observables at tree
level, then all such corrections from new particles are
loop suppressed and small. All of these allow a relatively low value
of $f$. According to the latest results from the $8~{\rm TeV}$ run
at the LHC, the constraints from Higgs couplings are by now
competing with electroweak precision tests and both combined exclude
$f$ up to $694~{\rm GeV}$ or $560~{\rm GeV}$ depending on the
implementation of the down-type Yukawa sector \cite{LHTlimits}.
With the new particles and a relatively
low value of $f$, rich phenomena of the LHT can be expected at the
LHC. Another interesting thing predicted in the LHT is that the
neutral and colorless weakly interacting stable $T$-odd $A_H$ can be
a good candidate for dark matter, which is also one of the problems
unsolved in current particle physics.

\par
The phenomenological study in the LHT is extensive. A study on
phenomenology of the LHT including effects of $T$-odd fermions at
tree level was reported in Ref.\cite{LHT-ph1}. The signals of
$T$-odd quarks in the LHT were analysed in Ref.\cite{LHT-ph2}. The
$W_H$ pair production and $W_H Z_H$ production at the LHC were
calculated up to the QCD NLO in Ref.\cite{DuSM} and
Ref.\cite{LHT-ph3-L.W}, respectively. Calculations on a $T$-odd gauge
boson production associated with a $T$-odd heavy quark at the LHC at the
QCD NLO were presented in Ref.\cite{LHT-ph4}. The production of
same-sign dileptons via SM mechanism at the LHC is rather rare,
therefore, it is a very helpful to choose the same-sign dilepton
production to search for the evidence of the LHT.
Both the ATLAS and CMS collaborations have adopted same-sign
dilepton events to search for new physics at the
$\sqrt{s}=7~{\rm TeV}$ LHC (LHC7) and the $\sqrt{s}=8~{\rm TeV}$
LHC (LHC8) \cite{CMS ssdl, ATLAS ssdl}.
By analyzing these signatures and other events from the ATLAS
at the LHC8, J. Reuter, {\it et al.} provided robust and
complementary LHT limits and constrained $f$ to be larger than
$638~{\rm GeV}$ \cite{LHTlimits}. In this work we study the
same-sign $T$-odd mirror quark pair production of
the first two generations followed by the subsequent decay
$q_- \to W_H q^{\prime} \to W A_H q^{\prime} \to
\ell \nu  A_H q^{\prime}$ up to the QCD NLO in the LHT at the
$\sqrt{s}=14~{\rm TeV}$ LHC (LHC14). The rest of this paper is organized as
follows: In Sec.II, we briefly review the related LHT theory.
The calculation strategy is presented in Sec.III.
The numerical results and discussions are provided in Sec.IV,
and finally a short summary is given in Sec.V.

\vskip 5mm
\section{Related LHT theory }\label{theory}
\par
In this section, we introduce briefly the LHT theory related to
our study. For the detailed LHT theory one can refer to the
literatures \cite{$T$-parity1, $T$-parity2, $T$-parity3,
LHTindetail}. The LHT is a nonlinear $\sigma$ model based on a
$SU(5)$ global symmetry, in which a subgroup $[SU(2)_1 \times
U(1)_1] \times [SU(2)_2 \times U(1)_2]$ is gauged.
At an energy scale of $f \sim 1~{\rm TeV}$, the $SU(5)$ global
symmetry spontaneously breaks down to its
$SO(5)$ subgroup. This symmetry breaking originates from the vacuum
expectation value (VEV) of an $SU(5)$ symmetric tensor field
$\Sigma$, written as
\begin{eqnarray}
\Sigma_0 = \langle \Sigma \rangle = \left(
 \begin{array}{ccccc}  & & 1_{2 \times 2} \\  & 1 & \\ 1_{2 \times 2} & & \end{array}
 \right).
\end{eqnarray}
This breaking gives rise to 14 Nambu-Goldstone (NG) bosons.
Simultaneously, the gauged symmetry $[SU(2)_1 \times
U(1)_1] \times [SU(2)_2 \times U(1)_2]$ breaks down to the SM $SU(2)_L
\times U(1)_Y$ symmetry. A ``pion'' matrix $\Pi$ is employed to describe the
14 NG bosons as
\begin{eqnarray}
\label{Pi}\addtolength{\arraycolsep}{3pt}\renewcommand{\arraystretch}{1.3}
 \Pi=\left(\begin{array}{ccccc}
-\frac{\omega^0}{2}-\frac{\eta}{\sqrt{20}} &
-\frac{\omega^+}{\sqrt{2}} &
  -i\frac{\pi^+}{\sqrt{2}} & -i\phi^{++} & -i\frac{\phi^+}{\sqrt{2}}\\
-\frac{\omega^-}{\sqrt{2}} &
\frac{\omega^0}{2}-\frac{\eta}{\sqrt{20}} &
\frac{v+h+i\pi^0}{2} & -i\frac{\phi^+}{\sqrt{2}} & \frac{-i\phi^0+\phi^P}{\sqrt{2}}\\
i\frac{\pi^-}{\sqrt{2}} & \frac{v+h-i\pi^0}{2} &\sqrt{4/5}\eta &
-i\frac{\pi^+}{\sqrt{2}} & \frac{v+h+i\pi^0}{2}\\
i\phi^{--} & i\frac{\phi^-}{\sqrt{2}} & i\frac{\pi^-}{\sqrt{2}} &
-\frac{\omega^0}{2}-\frac{\eta}{\sqrt{20}} & -\frac{\omega^-}{\sqrt{2}}\\
i\frac{\phi^-}{\sqrt{2}} &  \frac{i\phi^0+\phi^P}{\sqrt{2}} &
\frac{v+h-i\pi^0}{2} & -\frac{\omega^+}{\sqrt{2}} &
\frac{\omega^0}{2}-\frac{\eta}{\sqrt{20}}
\end{array}\right),\renewcommand{\arraystretch}{1.0}
\end{eqnarray}
where $(-i \pi^+/\sqrt{2},~ (v + h + i \pi^0)/2)^T$ is the
$T$-even $SU(2)$ Higgs doublet, identified as the SM Higgs
doublet, i.e., $h$ is the usual SM Higgs field, $v = 246~{\rm GeV}$
is the Higgs VEV, and $\pi^+$, $\pi^0$ are
Goldstone bosons eaten by the SM $W$, $Z$ bosons, respectively. The
fields $\eta$ and $\omega$ are additional Goldstone bosons, and $\Phi$
is a $T$-odd physical scalar triplet with components of
$\phi^{++}$, $\phi^+$, $\phi^0$ and $\phi^P$.

\par
From the symmetry breaking by the VEV $\Sigma_0$, the
$T$-odd gauge bosons $A_H$, $Z_H$ and $W_H$ get masses as
\begin{eqnarray}\label{mass-AH-VH}
 m_{A_H} \simeq \frac{1}{\sqrt{5}} g^{\prime} f
 \left( 1 - \frac{5}{8}\frac{v^2}{f^2} \right),~~~~
 m_{W_H} \simeq m_{Z_H} \simeq g f \left( 1 - \frac{1}{8}\frac{v^2}{f^2}
 \right).~~~~
\end{eqnarray}
The mass of the additional physical scalar triplet is given by
\begin{eqnarray}\label{mass-Higgs}
m_{\Phi} = \sqrt{2} m_h \frac{f}{v},
\end{eqnarray}
where $m_h$ is the mass of the SM Higgs boson, and all components of the
triplet $\Phi$ are degenerate at the
order of ${\cal O}\left(\frac{v^2}{f^2}\right)$. In principle, the
triplet $\Phi$ can contribute to our investigated subprocesses, but
the $\bar{q}_- \Phi q$ coupling strengths are of ${\cal O}
(\frac{v^2}{f^2})$ \cite{Feynman rules}. Then the contributions to
the amplitude involving such $\bar{q}_- \Phi q$ interactions
are suppressed by a factor of
$\frac{v^4}{f^4}$. Therefore, we can ignore the effects induced by
$\bar{q}_- \Phi q$ interactions in this paper.

\par
A consistent implementation of $T$-parity in the quark sector
requires the introduction of the $T$-odd mirror quarks for the SM quarks.
We denote the $T$-odd up- and down-type quarks as
$u^i_-$ and $d^i_-$ $(i = 1, 2, 3)$, respectively. Assuming a
universal and flavor independent Yukawa coupling $\kappa$, the
masses of the $T$-odd heavy quarks are given by
\begin{eqnarray}\label{m-Q}
 m_{u^i_-} \simeq \sqrt{2} \kappa f
 \left(
 1 - \frac{1}{8}\frac{v^2}{f^2}
 \right),~~~~
 m_{d^i_-} = \sqrt{2} \kappa f,~~~(i=1,2,3).
\end{eqnarray}

\par
The Feynman rules used in our calculations are listed in Table
\ref{Tab1} \cite{Feynman rules}. We note that the $T$-odd mirror
quark sector involves two Cabibbo-Kobayashi-Maskawa (CKM)-like
unitary mixing matrices $V_{Hu}$ and $V_{Hd}$, which satisfy
$V^{\dag}_{Hu}V_{Hd} = V_{CKM}$ \cite{Feynman rules}. In the
following calculations we take $V_{CKM}$ to be a unit matrix, then
we can take $V_{Hu}=V_{Hd}=I$.
\begin{table}[h]
\begin{center}
\begin{tabular}{|c|l||c|l|}
\hline
Vertex & ~~~~~~~~Feynman rule & Vertex & ~~~~~~~~~Feynman rule \\
\hline
&&& \\
$A_{H}^{\mu} \bar{u}_-^i u^j$ & $-i\Big(\frac{g'}{10}+\frac{g}{2}
\sin\theta_H\big)(V_{Hu})_{ij} \gamma^\mu P_L$ &
$A_{H}^{\mu} \bar{d}_-^i d^j$ & $i\Big(-\frac{g'}{10}+\frac{g}{2}
\sin\theta_H\Big)(V_{Hd})_{ij} \gamma^\mu P_L$\\
&&&\\
$Z_{H}^{\mu} \bar{u}_-^i u^j$ & $i\Big(\frac{g}{2}-\frac{g'}{10}
\sin\theta_H\Big)(V_{Hu})_{ij} \gamma^\mu P_L$ &
$Z_{H}^{\mu} \bar{d}_-^i d^j$ & $-i\Big(\frac{g}{2}+\frac{g'}{10}
\sin\theta_H\Big)(V_{Hd})_{ij} \gamma^\mu P_L$\\
&&&\\
$\omega^0 \bar{u}_-^i u^j$ & $\frac{g}{{2}m_{Z_H}} \Big[m^i_{u_-}
\Big(1+\frac{v^2}{8 f^2} -  $ &
 $\omega^0 \bar{d}_-^i d^j$ & $- \frac{g}{{2}m_{Z_H}}
\Big[m^i_{d_-}\Big(1-\frac{v^2}{4 f^2}+ $ \\
&&&\\
$$ & $\frac{\sin\theta_H}{\tan\theta_W}\Big)P_L
-m^j_u P_R \Big] (V_{Hu})_{ij}$ & &
 $\frac{\sin\theta_H}{\tan\theta_W}\Big)
P_L-m^j_d P_R\Big](V_{Hd})_{ij}$\\
&&&\\
$\eta \bar{u}_-^i u^j$ & $-\frac{g'}{{10}m_{A_H}}
\Big[m^i_{u_-}\Big(1+\frac{5v^2}{8f^2}+ $ &
$\eta \bar{d}_-^i d^j$ & $ - \frac{g'}{{10}m_{A_H}} \Big[m^i_{d_-}
\Big(1-\frac{5v^2}{4f^2}+ $  \\
&&&\\
$$&$\sin\theta_H\tan\theta_W\Big)P_L
-m^j_uP_R\Big](V_{Hu})_{ij}$&
$$&$\sin\theta_H\tan\theta_W\Big)
P_L -m^j_d P_R\Big](V_{Hd})_{ij}$\\
&&& \\
$\bar{q}_{-}^{\alpha} q_{-}^{\beta} G^{a}_{\mu}$ & $ig_s (T^a)_{\alpha\beta}\gamma^{\mu}$ &
$W^{+\mu} \bar{u}_{-}^i d_{-}^j$ & $\frac{ig}{\sqrt{2}} \delta_{ij} \gamma^{\mu}$ \\
&&& \\
\hline
\end{tabular}
\caption{\label{Tab1} The related LHT Feynman rules used in this
work, where $q_-=u_-,d_-,c_-,s_-$, $P_{L,R} =\frac{1}{2}\left(1 \mp \gamma_5 \right)$,
$i$ and $j$ are the generation
indices.}
\end{center}
\end{table}

\vskip 5mm
\section{Calculation strategy}\label{calc}
\par
We calculate the same-sign $T$-odd mirror quark pair production of the
first two generations at the LHC14 in the framework of the LHT up to
the QCD NLO, including the $T$-odd quark subsequent decays. We use
the developed FeynArts3.4 package \cite{FeynArts} to generate
Feynman diagrams and amplitudes under the t'Hooft Feynman gauge in
the LHT, and employ FormCalc5.4 program \cite{FormCalc} for
algebraic manipulation. We use the Passarino-Veltman (PV) method to
reduce a tensor integral to a linear combination of tensor
structures and scalar coefficients. We adopt the expressions in
Ref.\cite{ellis} to deal with the IR singularities in loop
integrals, and apply the expressions in Refs.\cite{hooft,
denner1991, denner2003} to implement the numerical evaluations for
the IR-safe $N$-point integrals.

\par
\subsection{LO cross section }
\par
\begin{figure}
\begin{center}
\includegraphics[width=0.8\textwidth]{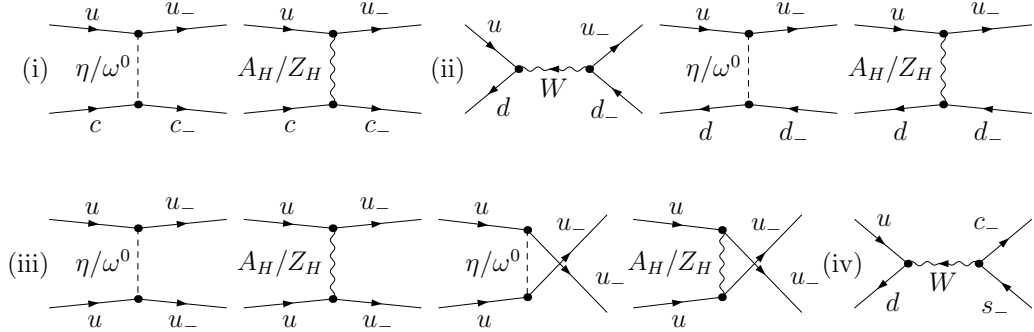} \\
\caption{ \label{treetype} The representative LO Feynman diagrams for the partonic
processes of type (i), (ii), (iii) and (iv), respectively.}
\end{center}
\end{figure}

The parton level processes, which contribute to the parent
process \ppqq with $q_-$ and $q^{\prime}_-$ same-sign charged, are
denoted as
\begin{eqnarray}
\label{subprocess} q(p_1)~+~q^{\prime}(p_2) \to
q_-(p_3)~+~q^{\prime}_-(p_4),
\end{eqnarray}
where $qq^{\prime} = uu, \bar{d}\bar{d}, cc, \bar{s}\bar{s}, uc,
\bar{d}\bar{s}, u\bar{d}, c\bar{s}, u\bar{s}, c\bar{d}$, and their
corresponding charge conjugations. $p_i~(i=1,2,3,4)$ stand for the four-momenta of
the incoming and outgoing particles. The LO partonic processes are
all induced by EW interactions, and can be divided into four types:
(i) The processes containing only $t$-channel Feynman diagrams. This type of partonic
processes include $uc \to u_{-}c_{-}$, $\bar{d}\bar{s} \to
\bar{d}_{-}\bar{s}_{-}$, $u\bar{s} \to u_{-}\bar{s}_{-}$, $c\bar{d}
\to c_{-}\bar{d}_{-}$ and their corresponding charge conjugations.
(ii) The processes involving $s$- and $t$-channel diagrams but no $u$-channel diagram.
(iii) The processes containing $u$- and $t$-channel diagrams
but no $s$-channel diagram. Each process of this type has identical two initial/final
particles, such as $uu \to u_{-}u_{-}$ partonic process. (iv) The processes
containing only a $s$-channel Feynman diagram via $W$-exchange, i.e.,
$u \bar{d} \to c_{-} \bar{s}_{-}$, $c \bar{s} \to u_{-} \bar{d}_{-}$
and their charge conjugations. We plot
representative Feynman diagrams for the four types of partonic processes
in Fig.\ref{treetype}.

\par
The LO cross section for the partonic process $qq^{\prime} \to
q_-q^{\prime}_-$ can be expressed as
\begin{eqnarray}
\hat{\sigma}_{LO}(\hat{s},q q^{\prime} \to q_-q^{\prime}_-)=
\frac{1} {1+\delta_{q_-q^{\prime}_-}} \frac{1}{4}\frac{1}{9}\frac{(2
\pi )^4}{2{\hat{s}}} \sum_{spin}\sum_{color} \int
|{\cal M}_{LO}|^2 d\Omega_2,
\end{eqnarray}
where $\frac{1}{4}$ and $\frac{1}{9}$ are from averaging
over the spins and colors of initial partons, and
$\frac{1}{1+\delta_{q_-q^{\prime}_-}}$ is due to the number of the
identical $T$-odd quarks in final state. $\sqrt{\hat{s}}$ is the
partonic center-of-mass (c.m.) colliding energy and ${\cal M}_{LO}$ is the LO
amplitude for the partonic process $qq^{\prime} \to
q_-q^{\prime}_-$. The two summations are taken over the spins and
colors of all the relevant initial and final particles, separately.
The integration is performed over the two-body phase space of the
final particles $q_-$ and $q^{\prime}_-$, and $d\Phi_2$ is the two-body
phase space element defined as
\begin{eqnarray}
d\Phi_2=\delta^{(4)} \left( p_1+p_2-p_3-p_4 \right) \prod_{i=3}^4
\frac{d^3 \vec{p}_i}{(2 \pi)^3 2 E_i}.
\end{eqnarray}

\par
The LO total cross section for the same-sign $T$-odd mirror quark
pair production of all the first two generations at the LHC,
denoted as $\sigma_{LO}(pp \to q_- q^{\prime}_-+X)$, can be obtained as
\begin{eqnarray}
\label{pp-total cross section}
&& \sigma_{LO}(pp \to q_- q^{\prime}_-+X) \nonumber \\
&& = \sum_{qq^{\prime}} \left\{
\int dx_A dx_B \Big[ G_{q/A}(x_A, \mu_f) G_{q^{\prime}/B}(x_B, \mu_f)
\hat{\sigma}_{LO}(x_{A}x_{B}s,q q^{\prime} \to q_- q^{\prime}_-, \mu_f) + (A
\leftrightarrow B) \Big] \right\},~~~~~~~
\end{eqnarray}
where the summation is taken over all the partonic processes of the same-sign
$T$-odd mirror quark pair production of the first two generations, i.e.,
$qq^{\prime} = uu,\bar{u}\bar{u}$, $\bar{d}\bar{d},dd$,
$cc,\bar{c}\bar{c}$, $\bar{s}\bar{s},ss$, $uc,\bar{u}\bar{c}$,
$\bar{d}\bar{s},ds$, $u\bar{d},\bar{u}d$, $c\bar{s},\bar{c}s$,
$u\bar{s},\bar{u}s$, $c\bar{d},\bar{c}d$.
$G_{i/P}~ (i = q,q^{\prime},P = A, B)$ is the parton distribution function (PDF)
of parton $i$ in proton $P$, $x_P$ is the momentum fraction of a parton in the proton $P$,
and $\mu_f$ is the factorization scale.

\par
\subsection{ NLO QCD corrections to \ppqq process}
\par
The NLO QCD corrections to the \ppqq process involve the following components:
(1) one-loop virtual corrections, (2) real
gluon/light-quark emission corrections, and (3) PDF counterterm contributions.
To regularize the ultraviolet (UV) and infrared (IR) divergences,
the dimensional regularization scheme in $D = 4 -2 \epsilon$ dimensions is
employed. According to the Kinoshita-Lee-Nauenberg (KLN) \cite{KLN}
theorem, the sum of the above three components gives a IR-finite result.

\par
The QCD one-loop level amplitudes for the partonic processes \qqqq
($qq^{\prime} = uu,\bar{u}\bar{u}$, $\bar{d}\bar{d},dd$,
$cc,\bar{c}\bar{c}$, $\bar{s}\bar{s},ss$, $uc,\bar{u}\bar{c}$,
$\bar{d}\bar{s},ds$, $u\bar{d},\bar{u}d$, $c\bar{s},\bar{c}s$,
$u\bar{s},\bar{u}s$, $c\bar{d},\bar{c}d$)
in the LHT are contributed by vertex and box Feynman diagrams.
Both UV and IR singularities arise in the amplitudes. The UV divergence
can be removed after performing the renormalization procedure.
Since these partonic processes are EW induced, only the wave functions
and masses of related colored
particles need to be renormalized here. We adopt the renormalization
constants defined in Ref.\cite{LHT-ph4} and use the on-shell
renormalization scheme. For the radiative processes, the phase space
with soft and collinear singularities are separated by adopting the
two cutoff phase space slicing (TCPSS) method \cite{TCPSS}, which is
intuitive, simple to implement, and relies on a minimum of process
dependent information in the NLO calculations. The real gluon
emission processes contain soft and collinear singularities, while
the real light-quark emission processes contain only collinear
singularities. The soft singularities in one-loop virtual corrections
can be canceled by those in the real gluon emission processes exactly.
The collinear singularities in the real gluon and light-quark emissions
are canceled by those in the virtual corrections and
corresponding PDF counterterms. The details about the PDF counterterms can be found in
Ref.\cite{TCPSS}.

\par
\subsection{ NLO QCD corrected decay widths of $T$-odd mirror quarks }
\par
We evaluate the impact of the terms of order $(\alpha_s/\pi)m_{V_H}^2/m_{q_-}^2$
$(V_H=Z_H,W_H,A_H)$ on the decays of $T$-odd mirror quarks, and find that their
contributions to the partial decay widths and branch ratios are below $0.2\%$ and
$0.01\%$, respectively. By neglecting the light quark masses and the terms of order
$(\alpha_s/\pi)m_{V_H}^2/m_{q_-}^2$, the NLO QCD corrected partial decay width for
the $q_{-} \to V_H q^{\prime}$ decay mode has the form as
\begin{eqnarray} \label{Width-1}
\Gamma_{NLO}(q_{-} \to V_H q^{\prime}) &=& \Gamma_{LO}(q_{-} \to V_H q^{\prime})
\left[1-\frac{2\alpha_s}{3\pi} \left(\frac{2\pi^2}{3}-\frac{5}{2}\right)\right], \nb \\
&&~~~~~~~~(q_-=u_-,d_-,c_-,s_-,\bar u_-,\bar d_-,\bar c_-,\bar s_-,~V_H=W_H,Z_H,A_H).
\end{eqnarray}
The explicit expressions for the LO partial decay widths of $T$-odd
up- and down-type mirror quarks, $\Gamma_{LO}(q_- \to V_H
q^{\prime})$, can be found in Appendix B of Ref.\cite{DuSM}.

\par
\subsection{Checks}
\par
The verification of the correctness of our calculations are made in the following
ways: Firstly, we employ the same
PDFs and input parameters as used in Ref.\cite{LHT-ph1} to calculate
the LO cross section. The numerical results we obtained are in good
agreement with those shown in Fig.6 of Ref.\cite{LHT-ph1}.
Secondly, we check the cancelations of UV and IR divergences
both analytically and numerically. Finally, we check the independence
of the NLO QCD corrected total cross section on the soft cutoff $\delta_s$
in the range of $5 \times 10^{-6} < \delta_s < 5 \times 10^{-3}$
with $\delta_c = \delta_s /100$.
That is a good indirect check for the correctness of our
evaluations. In the further numerical calculations, we set $\delta_s
= 5 \times 10^{-5}$ and $\delta_c = 5 \times 10^{-7}$.

\vskip 5mm
\section{Numerical results and discussions}\label{numres}
\par
\subsection{Input parameters}
\par
In this section we present and discuss the numerical results for
the same-sign $T$-odd mirror quark pair production of the
first two generations at both the LO and the QCD NLO.
The SM input parameters for our calculations are taken as \cite{PDG2012}
\begin{eqnarray}
\alpha^{-1}_{ew} = 137.036,~~m_W = 80.385~{\rm GeV},
~~ m_Z = 91.1876~{\rm GeV}.
\end{eqnarray}
The Weinberg angle is fixed in the on-shell scheme as
$\sin^2\theta_W = 1 - ({\frac{m_W}{m_Z}})^2 = 0.2229$.
We adopt the CTEQ6L1 PDFs and CTEQ6.6 PDFs
\cite{cteq6} for the LO and QCD NLO calculations, separately.
We take one-loop and two-loop running $\alpha_{s}$ in the LO and QCD NLO
calculations \cite{PDG2012}, and the strong coupling
constant $\alpha_s(\mu)$ is determined by the QCD parameter
$\Lambda_5^{LO} = 165~{\rm MeV}$ for the CTEQ6L1 at the LO and
$\Lambda_5^{\overline{MS}} = 226~{\rm MeV}$ for the CTEQ6.6 at the QCD NLO
\cite{PDG2012}, respectively.
The factorization and the renormalization scales are set to be equal
($\mu_f = \mu_ r = \mu$) for simplicity, and the central scale value is
defined as $\mu_0 = (m_{u_-} + m_{d_-})$. The masses for the light
quarks ($u$, $d$, $c$, $s$) are set to be zero in our
numerical calculations.

\par
\subsection{Integrated cross sections}
\par
\begin{figure}
\begin{center}
\includegraphics[width=0.48\textwidth]{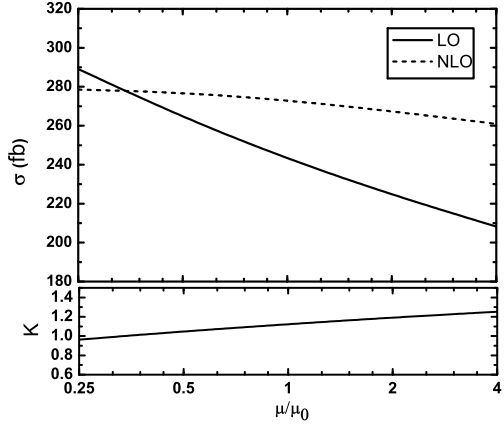}
\caption{ \label{mu} The dependence of the LO, NLO QCD corrected
total cross sections and the QCD $K$-factor on the factorization/renormalization
scale for the \ppqq process at the LHC14. }
\end{center}
\end{figure}

In Fig.\ref{mu} we show the dependence of the LO, NLO QCD corrected
total cross sections and the QCD $K$-factor on the factorization/renormalization
scale for the same-sign $T$-odd mirror quark pair production of all the
first two generations at the LHC14. The LHT parameters are taken as
$f = 700~{\rm GeV}$ and $\kappa$ = 1. From this figure we can see
that at the central scale $\mu_0$ the LO and NLO QCD corrected
total cross sections are $243.0~fb$ and $273.1~fb$, separately,
and the corresponding $K$-factor ($K \equiv \frac{\sigma_{NLO}}
{\sigma_{LO}}$) is $1.12$. If we define the upper and lower relative
scale uncertainties as
\begin{eqnarray}
\eta_{{\rm upper}} = \frac{{\rm max}
\left[ \sigma(\mu) - \sigma(\mu_0) \right]}{\sigma(\mu_0)},~~~~
\eta_{{\rm lower}} = \frac{{\rm min}
\left[ \sigma(\mu) - \sigma(\mu_0) \right]}{\sigma(\mu_0)},~~~~~~~~~~ \mu \in [\mu_0/4,~ 4 \mu_0],
\end{eqnarray}
the relative scale uncertainties of the LO and NLO total cross sections
are $\left(^{+19.0\%}_{-14.3\%}\right)$ and
$\left(^{+2.0\%}_{-4.5\%}\right)$, respectively.
We see from Fig.\ref{mu} and above data that the NLO QCD correction reduces
the scale uncertainty significantly.

\par
PDF is another source of theoretical uncertainty. In this work
we consider the PDF uncertainty at the QCD NLO at the central
scale $\mu_0$. For a given parametrization of the PDFs, the PDF
uncertainty comes from the experimental uncertainties on the fitted
data. We employ the definition of PDF uncertainty on the hadronic
cross section given in Ref.\cite{pdf uncertainty}. By making use of
the 44 additional PDF sets provided by the CTEQ6.6, we obtain the
NLO QCD corrected integrated cross section with the PDF uncertainty
at $\mu=\mu_0$ as $\sigma_{NLO}=273.1^{+9.8}_{-8.3}~fb$. The
corresponding relative PDF uncertainties are
$\left(^{+3.6\%}_{-3.0\%}\right)$, which are of the same level as
the relative scale uncertainties at the QCD NLO. By adding linearly
the scale and PDF uncertainties, the NLO integrated cross section at $\mu=\mu_0$ is
$\sigma_{NLO}=273.1^{+~5.4~+9.8}_{-12.2~-8.3}~fb=273.1^{+15.2}_{-20.5}~fb$.
Although we consider both the scale uncertainty and PDF uncertainty
for the NLO QCD corrected integrated cross section, the combined
uncertainty is still much smaller than the LO scale uncertainty,
which has the value of $\sigma_{LO}=243.0^{+46.1}_{-34.7}~fb$ at
$\mu=\mu_0$.

\par
\begin{figure}
\begin{center}
\includegraphics[width=0.48\textwidth]{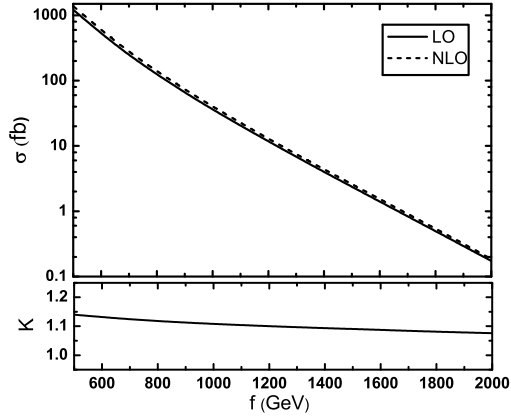}
\caption{ \label{f} The dependence of the LO, NLO QCD corrected
total cross sections and the QCD $K$-factor on the
global symmetry breaking scale $f$ for the \ppqq
process at the LHC14.}
\end{center}
\end{figure}

The dependence of the LO, NLO QCD corrected total cross sections and
the QCD $K$-factor on the global symmetry breaking scale $f$ are depicted
in Fig.\ref{f}, with $\kappa=1$ and $f$ varying from $500~{\rm GeV}$ to $2~{\rm TeV}$.
The figure shows that the LO integrated cross section is $1213.2~fb$ at
$f = 500~{\rm GeV}$, and decreases rapidly to $0.173~fb$ at $f= 2~{\rm TeV}$.
The NLO QCD corrected integrated cross section has the same tendency as the LO
cross section, having the values of $\sigma_{NLO}=1382.8~fb$ at
$f = 500~{\rm GeV}$ and $\sigma_{NLO}=0.186~fb$ at $f = 2~{\rm TeV}$.
This sensitive dependence of both the LO and NLO QCD corrected integrated
cross sections for the \ppqq process on the parameter $f$ is clearly
displayed in Fig.\ref{f}. It can be explained by that with the increment of
the global symmetry breaking scale $f$, the masses of $T$-odd mirror quarks
become heavier, which suppresses the phase space of final state.
The QCD $K$-factor varies from $1.14$ to $1.08$ as the increment of $f$ from
$500~{\rm GeV}$ to $2~{\rm TeV}$, and exhibits weak dependence on the
global symmetry breaking scale $f$.

\par
\begin{figure}
\begin{center}
\includegraphics[width=0.48\textwidth]{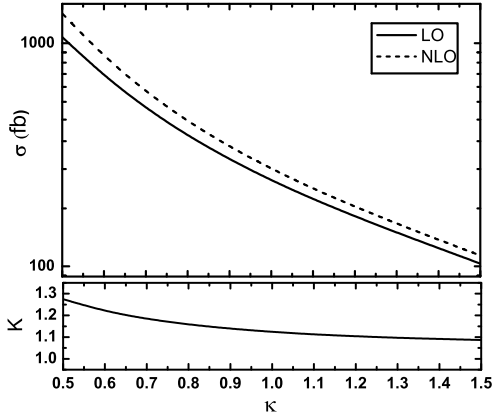}
\caption{ \label{kappa} The dependence of the LO, NLO QCD corrected
total cross sections and the QCD $K$-factor on the
$T$-odd mirror quark Yukawa coupling $\kappa$ for the \ppqq
process at the LHC14.}
\end{center}
\end{figure}

In Fig.\ref{kappa} we present the LO, NLO QCD corrected total cross sections
and the QCD $K$-factor for the same-sign $T$-odd mirror quark pair production
of all the first two generations at the LHC14 as functions of the $T$-odd mirror
quark Yukawa coupling $\kappa$ by taking $f = 700~{\rm GeV}$.
A similar behavior as depicted in Fig.\ref{f} can be found in Fig.\ref{kappa}.
It is because the $T$-odd mirror quark mass is simultaneously
proportional to the parameters $\kappa$ and $f$ as shown in
Eq.(\ref{m-Q}). The LO integrated cross sections are $1064.4~fb$
and $102.8~fb$, while the NLO QCD corrected integrated cross sections are
$1357.1~fb$ and $111.7~fb$, at $\kappa = 0.5$ and $\kappa = 1.5$, respectively.
The corresponding $K$-factor decreases from $1.27$ to $1.09$ when $\kappa$
varies from $0.5$ to $1.5$, which shows that the $K$-factor is more
sensitive to the Yukawa coupling $\kappa$ than to the global symmetry breaking
scale $f$.

\par
\subsection{ Distributions of final products}
\par
\begin{figure}
\begin{center}
\includegraphics[width=0.48\textwidth]{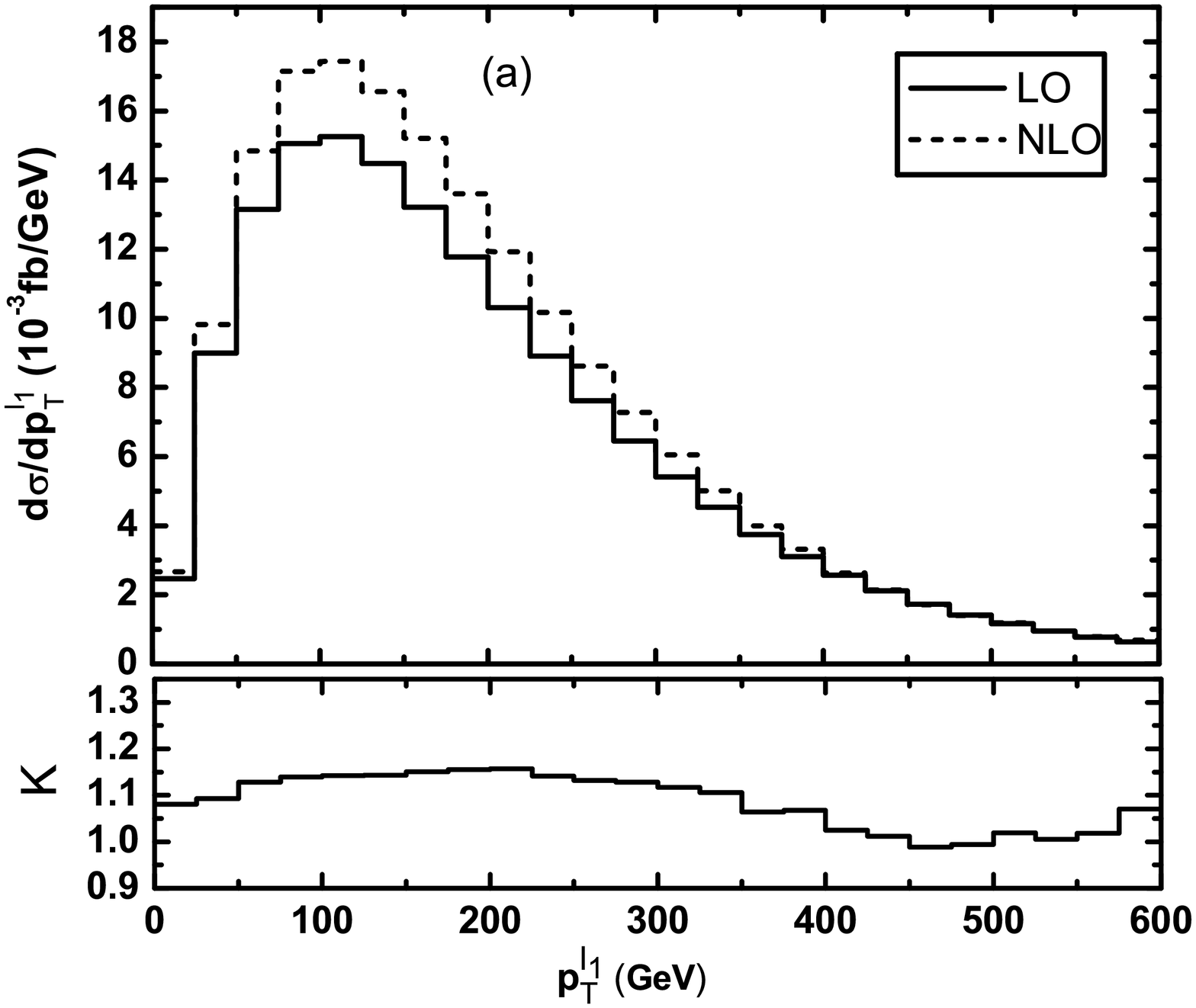}
\includegraphics[width=0.48\textwidth]{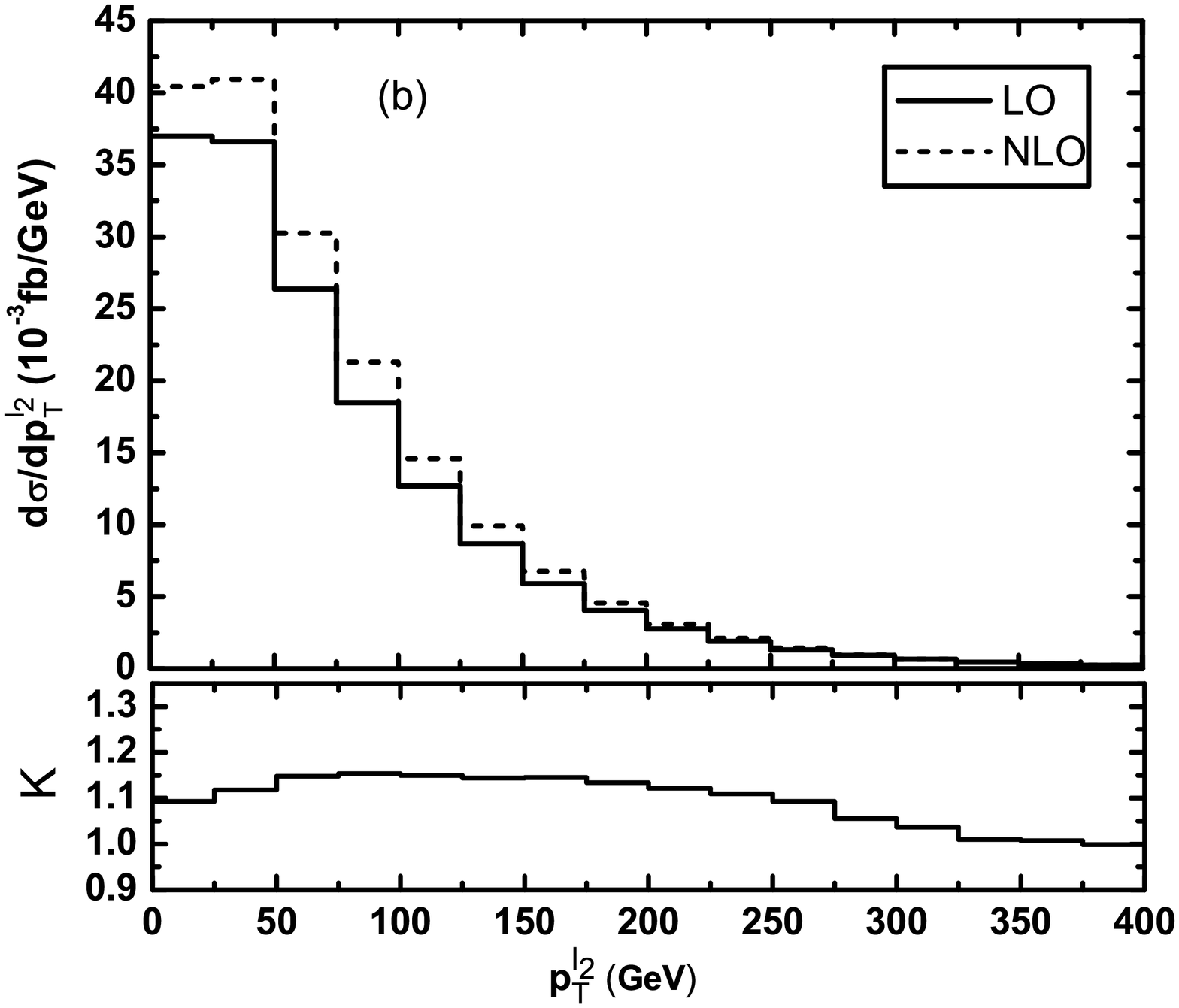}
\includegraphics[width=0.48\textwidth]{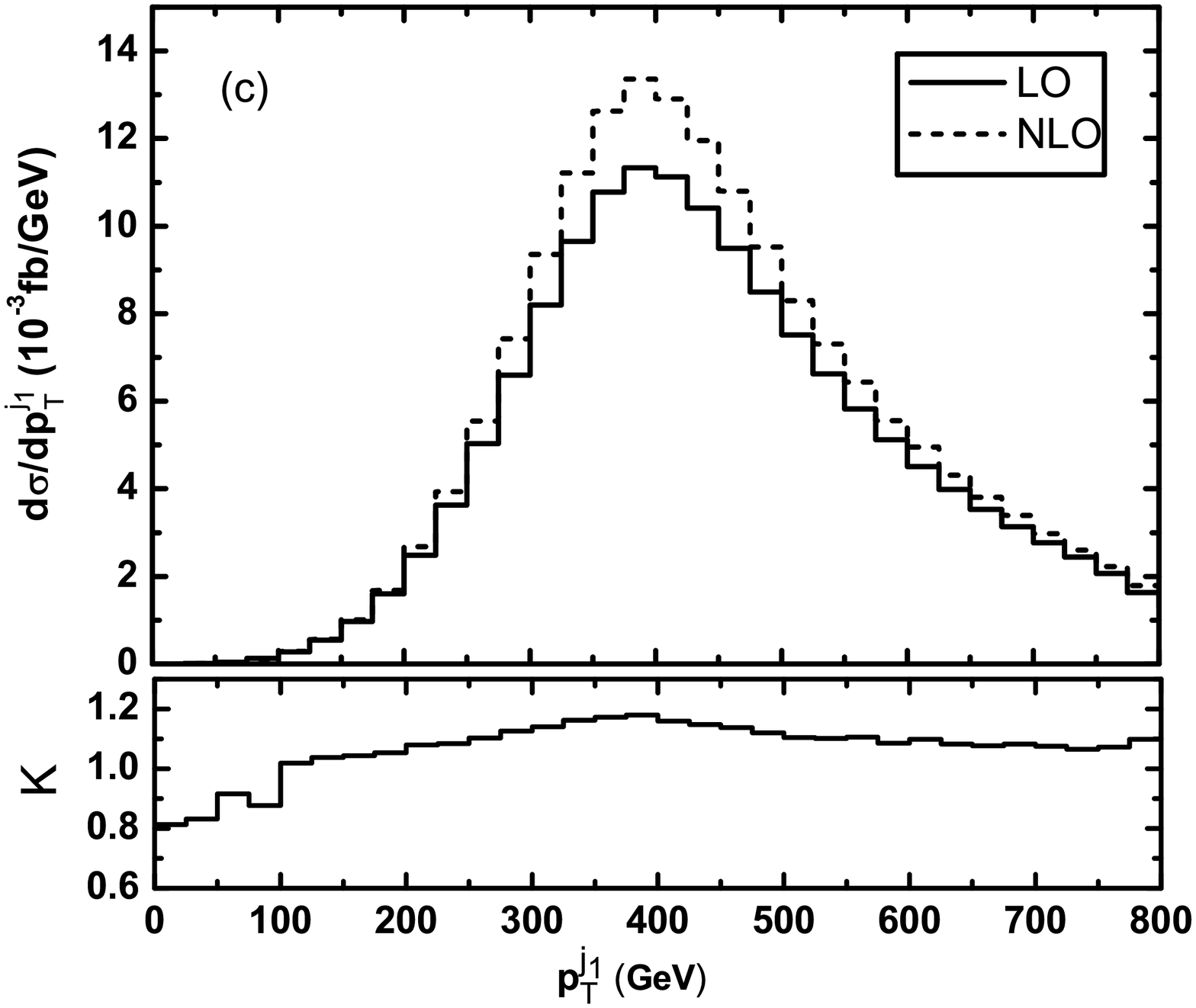}
\caption{ \label{pt} The LO, NLO QCD corrected transverse momentum
distributions and corresponding $K$-factors for the $pp \to
q_-q^{\prime}_-  \to \ell^{\pm} \ell^{\prime \pm} + jets +
\slashed{E}_T+X$ process at the LHC14. (a) leading lepton
$\ell_1$, (b) second lepton $\ell_2$, (c) leading jet $j_1$.}
\end{center}
\end{figure}
\begin{figure}
\begin{center}
\includegraphics[width=0.48\textwidth]{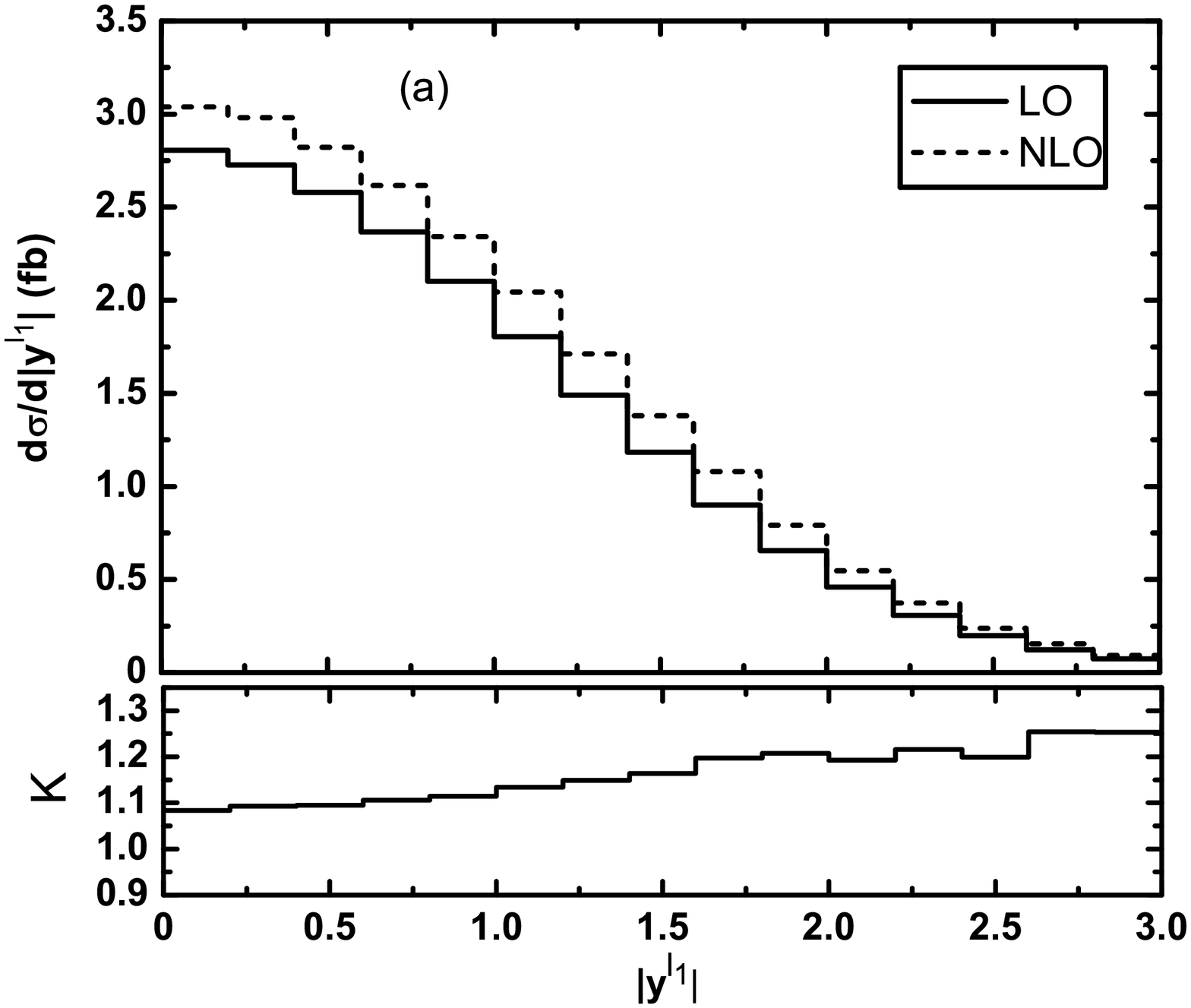}
\includegraphics[width=0.48\textwidth]{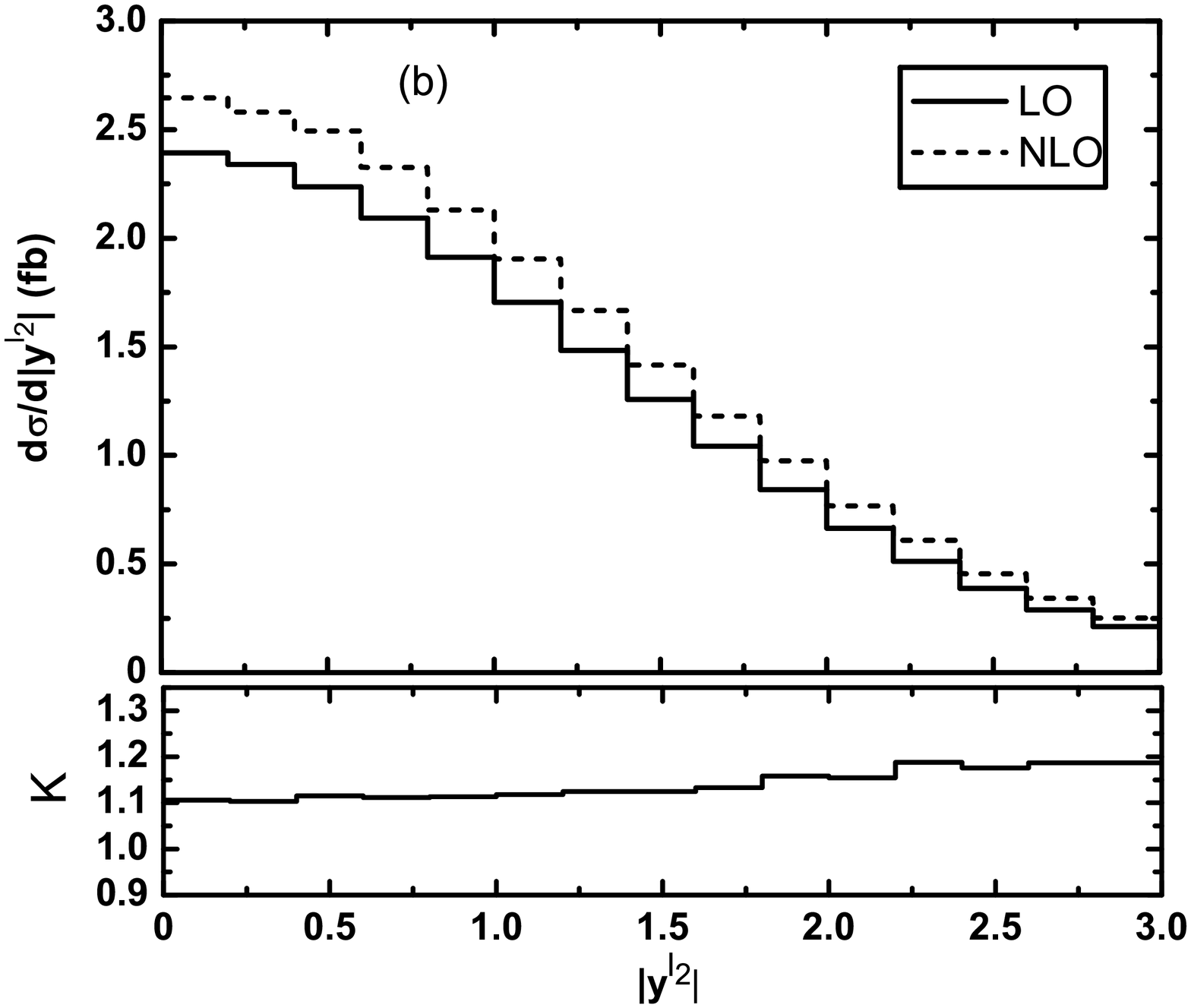}
\includegraphics[width=0.48\textwidth]{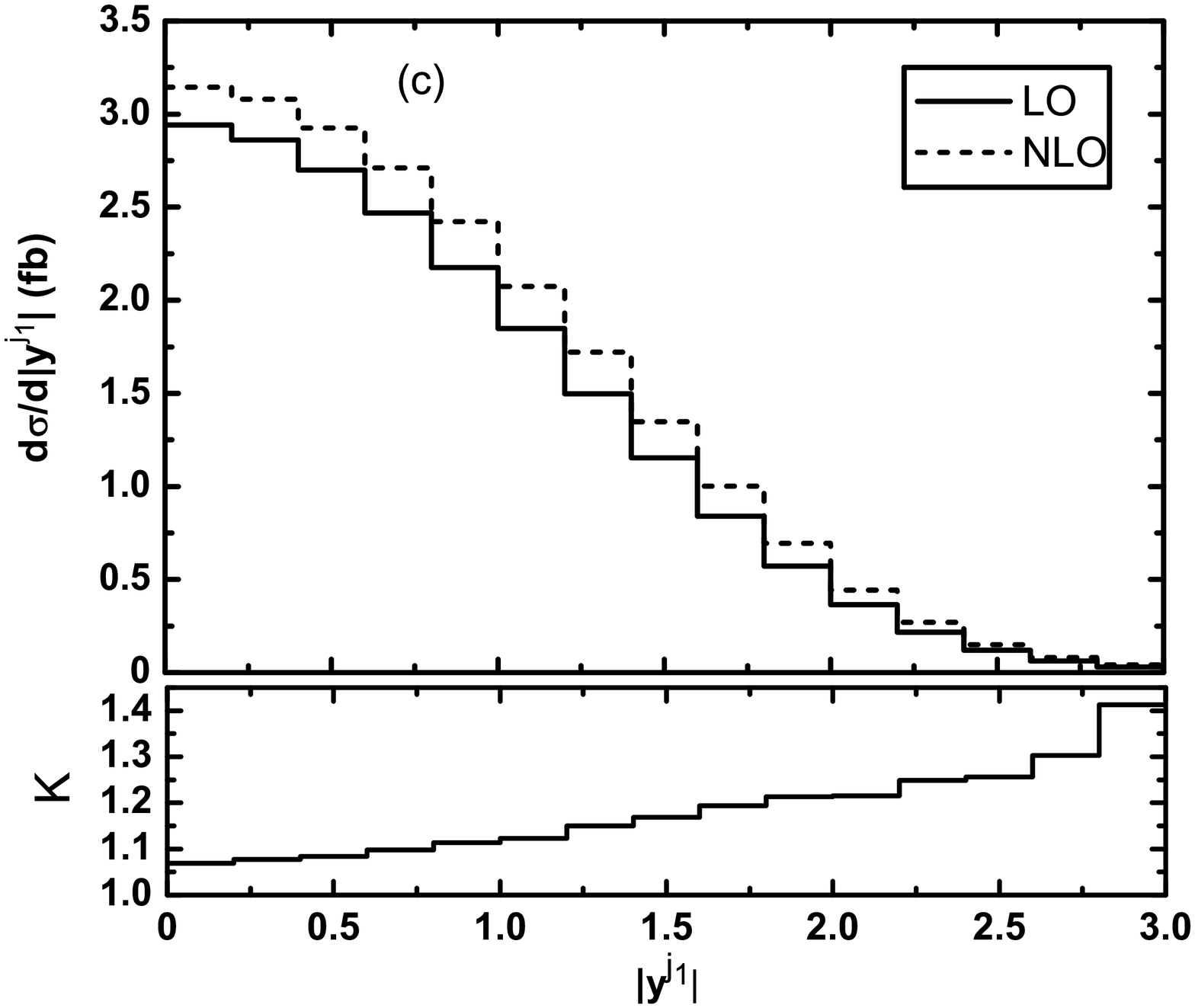}
\caption{ \label{yt}  The LO, NLO QCD corrected rapidity
distributions and corresponding $K$-factors for the $pp \to
q_-q^{\prime}_-  \to \ell^{\pm} \ell^{\prime \pm} + jets +
\slashed{E}_T+X$ process at the LHC14. (a) leading lepton
$\ell_1$, (b) second lepton $\ell_2$, (c) leading jet $j_1$. }
\end{center}
\end{figure}
\begin{figure}
\begin{center}
\includegraphics[width=0.48\textwidth]{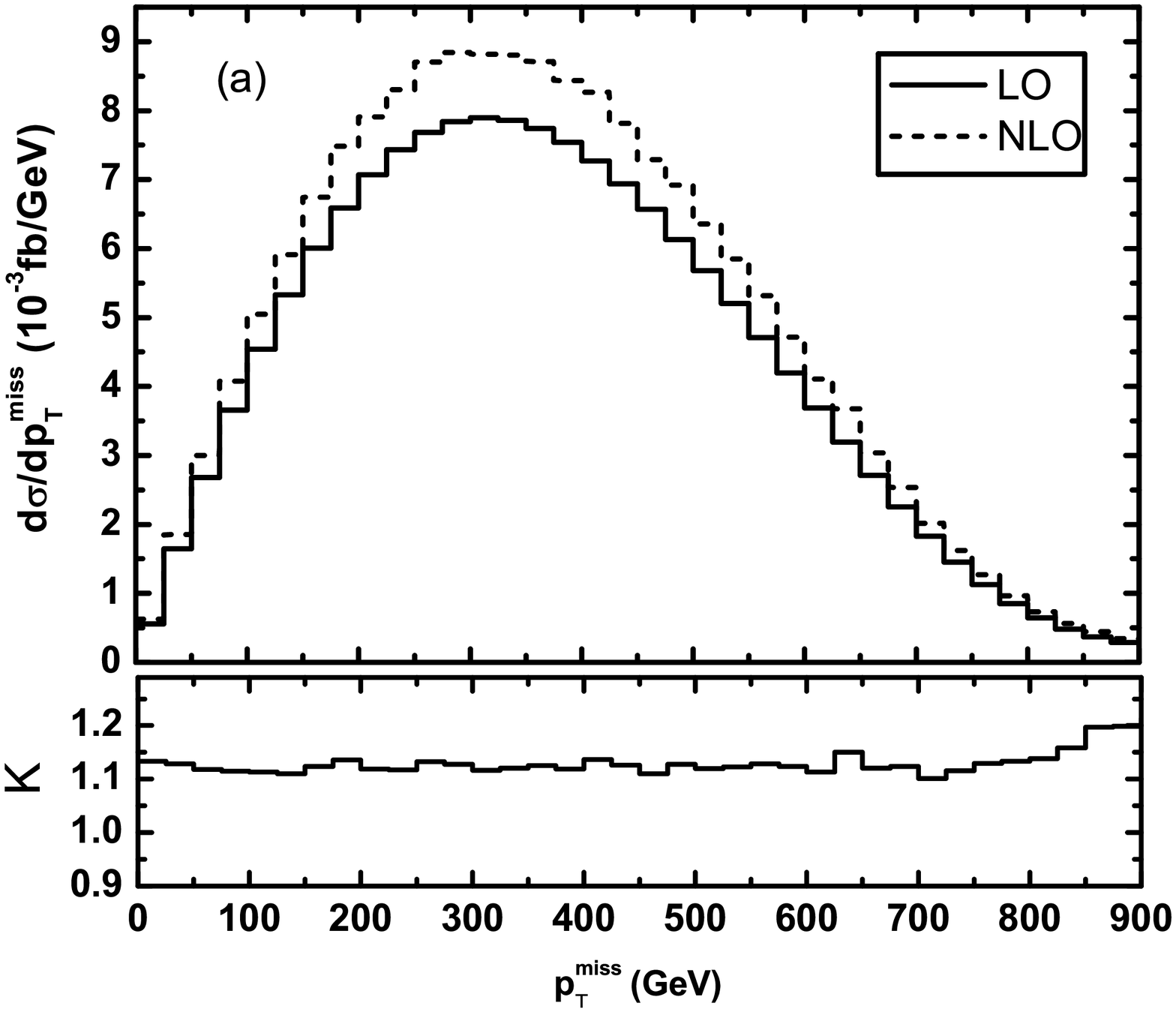}
\includegraphics[width=0.48\textwidth]{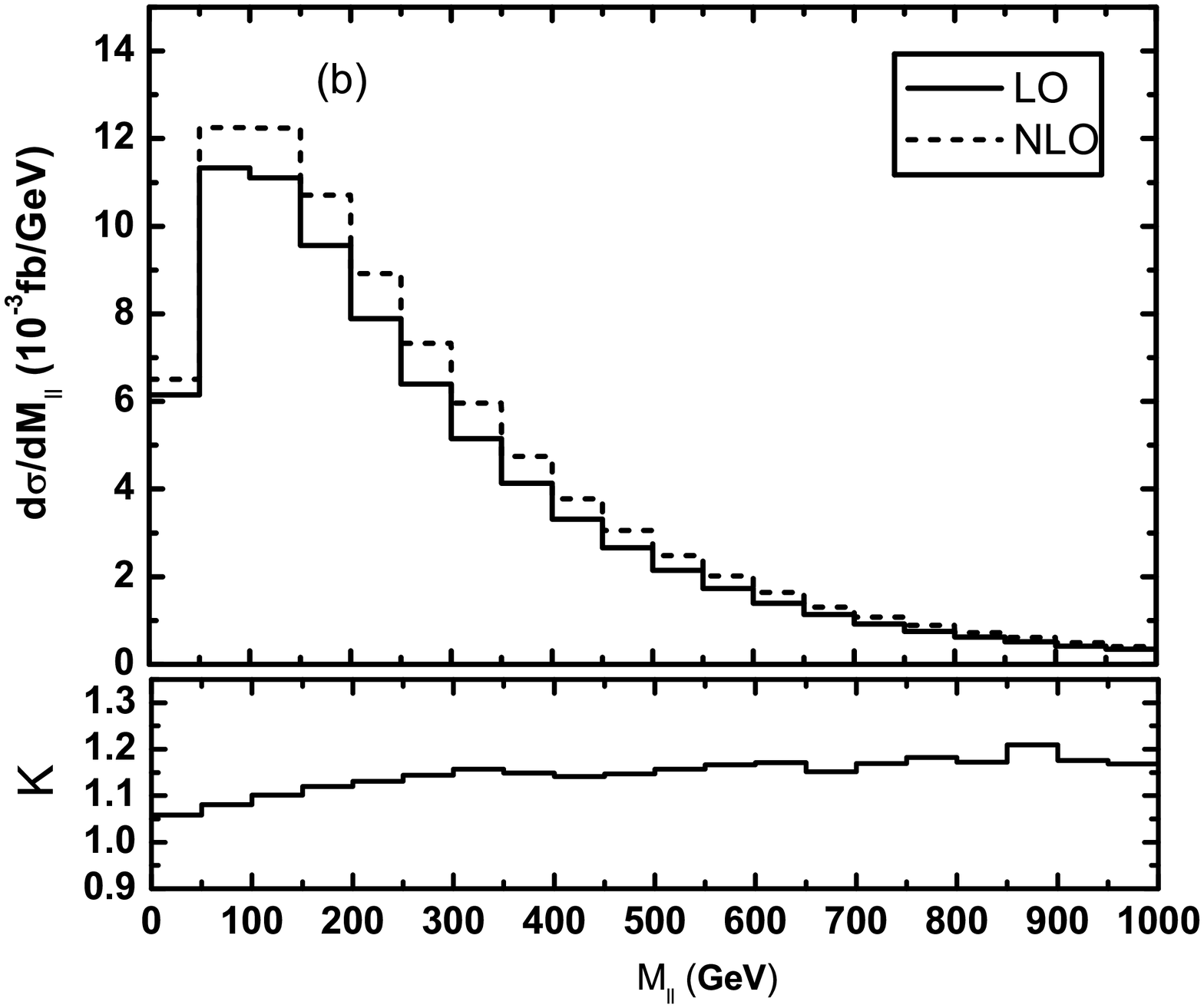}
\caption{ \label{Mll_ET} The LO and NLO QCD corrected distributions of the
(a) missing transverse momentum $p_T^{{\rm miss}}$ and (b) invariant mass
of same-sign lepton pair $M_{\ell\ell}$, for the
$pp \to q_-q^{\prime}_- \to \ell^{\pm} \ell^{\prime \pm} + jets + \slashed{E}_T + X$
process at the LHC14.}
\end{center}
\end{figure}
\begin{figure}
\begin{center}
\includegraphics[width=0.48\textwidth]{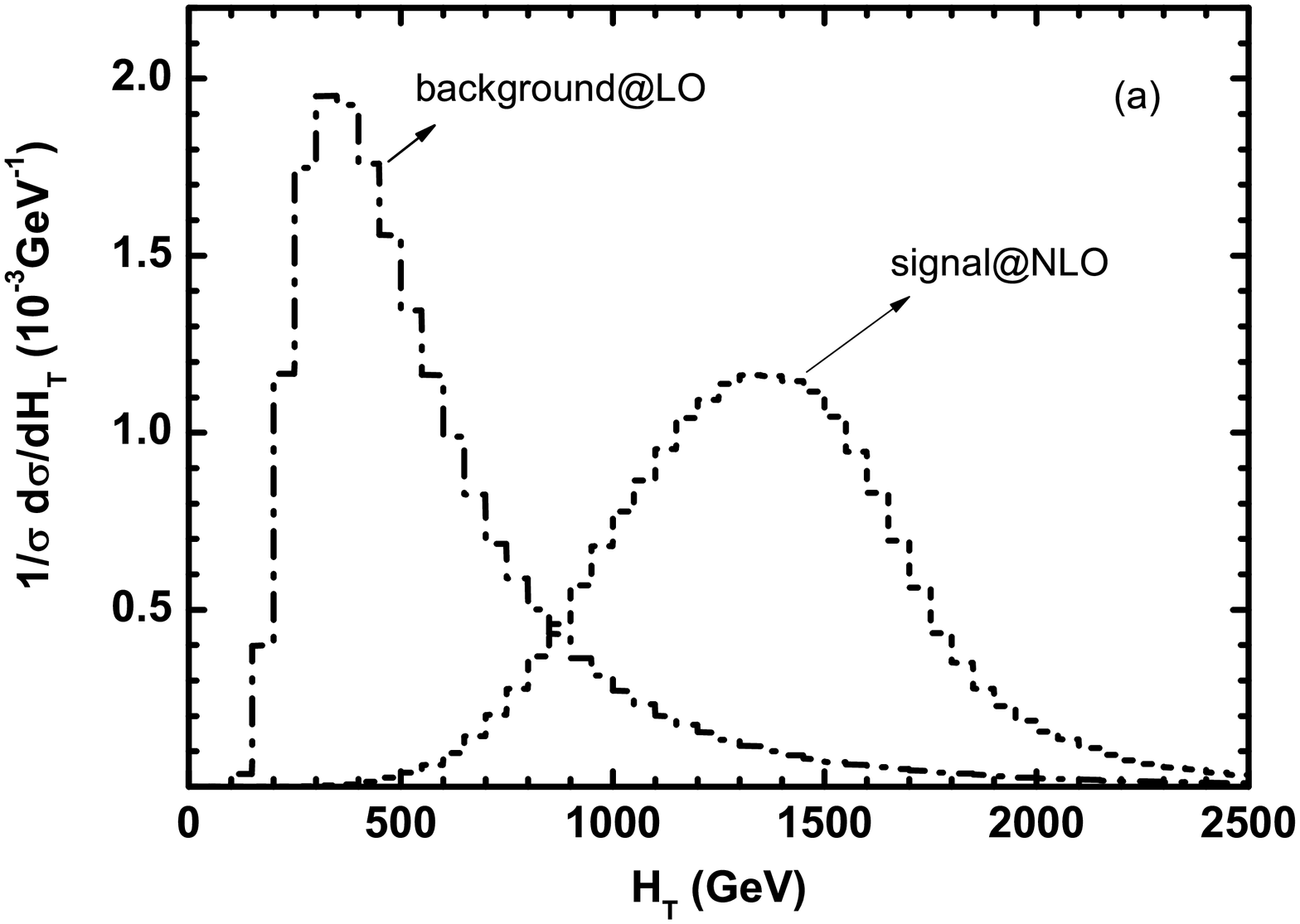}
\includegraphics[width=0.48\textwidth]{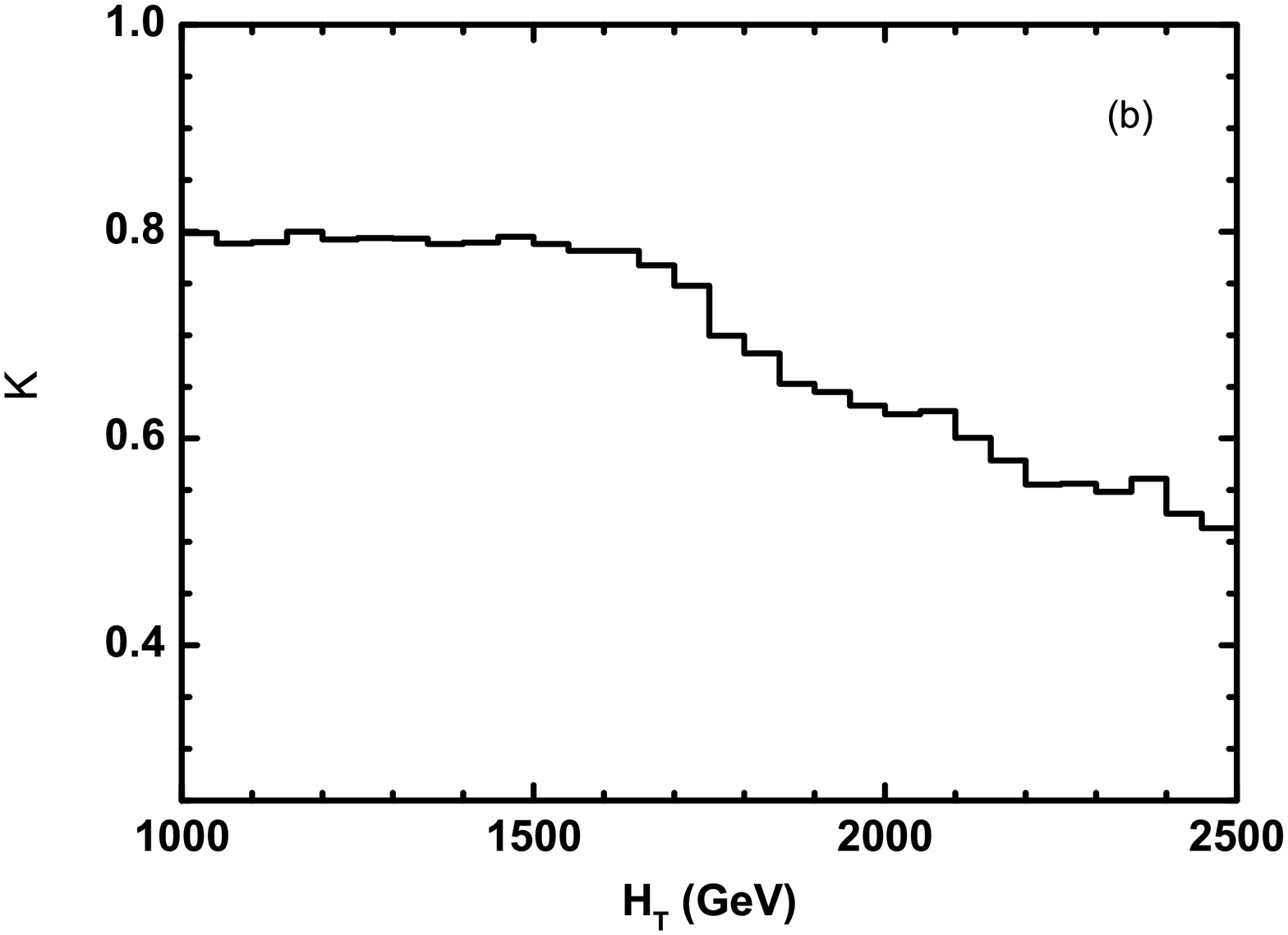}
\caption{ \label{HT} (a) The normalized $H_T$ distributions for the signal process
$pp \to q_-q^{\prime}_- \to \ell^{\pm} \ell^{\prime \pm} + 2~jets + \slashed{E}_T + X$
at the QCD NLO and the SM background process
$pp \to W^{\pm}W^{\pm} q q^{\prime} \to \ell^{\pm} \ell^{\prime \pm} + 2~jets + \slashed{E}_T + X$ at the LO at the LHC14.
(b) The QCD $K$-factor of the $H_T$ distribution for the signal process. }
\end{center}
\end{figure}

From Eqs.(\ref{mass-AH-VH}) and (\ref{m-Q}) we know that we have the mass
spectrum with $m_{q_-} > m_{W_H} > m_{A_H}$ and $A_H$ being the lightest
$T$-odd particle by setting $\kappa > 0.45$ \cite{LHTlimits}. In the following
we take $\mu=\mu_0$, $f = 700~{\rm GeV}$ and $\kappa = 1$ only for demonstration.
In this case we get $m_{u_-} = m_{c_-} = 974.67~{\rm GeV}$,
$m_{d_-} = m_{s_-} = 989.95~{\rm GeV}$, $m_{W_H} = m_{Z_H} = 442.06~{\rm GeV}$
and $m_{A_H} = 99.24~{\rm GeV}$. Then the main decay modes of the $T$-odd
mirror quark $q_-$ are $q_- \to W_H q^{\prime}$, $q_- \to Z_H q$ and $q_- \to A_H q$.
We assume that the total decay width of $q_-$ is the summation of the partial
decay widths of these three decay modes, i.e., $\Gamma_{q_-}^{{\rm tot}} =
\Gamma(q_- \to W_H q^{\prime}) + \Gamma(q_- \to Z_H q) + \Gamma(q_- \to A_H q)$.
By adopting the decay width expressions in Eq.(\ref{Width-1}), we obtain
the branch ratios for these decay channels as
\begin{eqnarray}
&& Br(U_- \to W_H D,~ Z_H U,~ A_H U)=59.84\%,~ 29.80\%,~ 10.36\%, \nonumber \\
&& Br(D_- \to W_H U,~ Z_H D,~ A_H D)=62.87\%,~ 31.55\%,~ 5.58\%,~~~~~~~~~~
\end{eqnarray}
where $U$ and $D$ are the up- and down-type SM quarks of the first two generations,
and $U_-$ and $D_-$ are the corresponding $T$-odd mirror quarks.
Via the following cascade decays of the $T$-odd mirror quarks,
\begin{eqnarray}\label{decay-chain}
&& U_- \to W_H D \to W A_H D \to \ell \nu A_H D, \nonumber \\
&& D_- \to W_H U \to W A_H U \to \ell \nu A_H U,~~~~~~
(\ell = e^{\pm}, \mu^{\pm},~~~ \nu = \nu_e, \bar\nu_e, \nu_{\mu}, \bar\nu_{\mu}),
\end{eqnarray}
the same-sign $T$-odd mirror quark pair production gives rise to
the interesting same-sign dilepton signature associated with jets and
a significant amount of missing energy,
\begin{eqnarray}
\label{lpm-lpm-signal}
~~~~
p p \to q_- q^{\prime}_-  \to \ell^{\pm} \ell^{\prime \pm} + jets + \slashed{E}_T + X,~~~~~
(\ell^{\pm}\ell^{\prime \pm} = e^{\pm}e^{\pm}, \mu^{\pm}\mu^{\pm}, e^{\pm}\mu^{\pm}).
\end{eqnarray}
In our numerical calculations we take
$Br(W \to e \nu_e) = 10.75\%$ and $Br(W \to \mu \nu_{\mu})= 10.57\%$ \cite{PDG2012}.
All the kinematic distributions for the same-sign dilepton signal process (\ref{lpm-lpm-signal})
 investigated in our paper are summed over $e^{\pm}e^{\pm}$, $\mu^{\pm}\mu^{\pm}$ and $e^{\pm}\mu^{\pm}$
final states.\footnote{ Besides the kinematic distributions for the signal process in Figs.\ref{pt}-\ref{HT},
the $H_T$ distribution for the SM background process
$pp \to W^{\pm}W^{\pm} q q^{\prime} \to \ell^{\pm} \ell^{\prime \pm} + 2~jets + \slashed{E}_T + X$ in Fig.\ref{HT}(a)
is also summed over $e^{\pm}e^{\pm}$, $\mu^{\pm}\mu^{\pm}$ and $e^{\pm}\mu^{\pm}$ final states.}

\par
Among the two same-sign leptons in final state, the leading lepton $\ell_1$
and the second lepton $\ell_2$ are defined as
\begin{eqnarray}
p_T^{\ell_1} > p_T^{\ell_2}.
\end{eqnarray}
For the jets originating from the decays of the produced same-sign $T$-odd mirror
quarks and the real gluon/light-quark emissions, we merge the proto-jets to
define the final jets by adopting the Cambrige/Aachen (C/A) jet algorithm
\cite{jet algorithm}, setting $R = 0.7$. After performing the jet merging
procedure, there are three kinds of same-sign dilepton events at the QCD NLO,
i.e., $\ell^{\pm} \ell^{\prime \pm} + 1~jet + \slashed{E}_T$,
$\ell^{\pm} \ell^{\prime \pm} + 2~jets + \slashed{E}_T$ and
$\ell^{\pm} \ell^{\prime \pm} + 3~jets + \slashed{E}_T$,
for the same-sign $T$-odd mirror quark pair production.
We define the jet with the largest transverse momentum in the two-jet or
three-jet event as well as the only jet in the one-jet event as the leading
jet, denoted as $j_1$. The final produced two heavy photons and two neutrinos can
not be detected directly, and are manifested as the missing of transverse energy.
We use the Monte Carlo method and adopt the narrow width approximation (NWA)
to generate these signal events, as well as the background events used in
the $H_T$ distribution for the SM background (``background@LO'' histogram in Fig.\ref{HT}(a)).

\par
The LO, NLO QCD corrected transverse momentum distributions and corresponding
$K$-factors of the leading lepton $\ell_1$, second lepton $\ell_2$
and leading jet $j_1$ are shown in Figs.\ref{pt}(a), (b) and (c), respectively.
In the range of $p_T^{\ell_1} \in [50,~350]~{\rm GeV}$, the NLO QCD correction
enhances the LO $p_T^{\ell_1}$ distribution by more than $10\%$,
and the $K$-factor is about $1.11 - 1.16$. Both the LO and NLO QCD corrected
$p_T^{\ell_1}$ distributions reach their maxima at the position of
$p_T^{\ell_1} \sim 100~{\rm GeV}$ with $K \sim 1.14$.
The $p_T$ distributions of the second lepton
are quite different from the corresponding distributions of the leading lepton.
Both the LO and NLO QCD corrected $p_T^{\ell_2}$ distributions peak at the
position of $p_T^{\ell_2} \sim 25~{\rm GeV}$, while
the $K$-factor of the $p_T^{\ell_2}$ distribution seems to be similar with that of
the $p_T^{\ell_1}$ distribution. The NLO QCD correction also enhances the LO
transverse momentum distribution of the leading jet for $p_T^{j_1} > 100~ {\rm GeV}$,
and the $K$-factor is about $1.09 - 1.18$ in the range of $p_T^{j_1} \in [250,~600]~{\rm GeV}$.
The peaks of the LO and NLO QCD corrected $p_T^{j_1}$ distributions are located
at $p_T^{j_1} \sim 375~{\rm GeV}$ with $K \sim 1.18$.
We also see from these figures that the NLO QCD correction is less than $18\%$
for the $p_T$ distributions of the leading lepton, second lepton and leading jet
in the plotted $p_T$ regions.

\par
In Figs.\ref{yt}(a), (b) and (c) we depict the LO and NLO QCD corrected
rapidity distributions of the two same-sign leptons and leading jet, separately.
We can see that all these rapidity distributions have similar behavior.
The final two same-sign leptons and leading jet prefer to be produced in
the central rapidity region. The $K$-factors in the three figures increase
with the increment of $|y|$. For the leading jet rapidity distribution,
the $K$-factor changes from $1.07$ at $|y^{j_1}|=0$ to $1.41$ at $|y^{j_1}|=3$.

\par
The LO and NLO QCD corrected distributions of the missing transverse momentum
and invariant mass of two same-sign leptons in final state are presented in
Figs.\ref{Mll_ET}(a) and (b), respectively. The missing transverse momentum
is carried by the final two heavy photons and two neutrinos originating from
the cascade decays of $T$-odd mirror quarks shown in Eq.(\ref{decay-chain}).
As shown in the two figures, both the LO and NLO QCD corrected missing
transverse momentum distributions reach their maxima at
$p^{{\rm miss}}_T \sim 275~ {\rm GeV}$ with $K \sim 1.13$, and
the LO and NLO QCD corrected invariant mass distributions of the same-sign
lepton pair peak at $M_{\ell\ell} \sim 50~{\rm GeV}$ with $K=1.08$.

\par
In the following discussion we only consider the two-jet event,
$\ell^{\pm} \ell^{\prime \pm} + 2~jets + \slashed{E}_T$, for the same-sign
$T$-odd mirror quark pair production of the first two generations at the LHC.
Then the SM background mainly comes from the same-sign $W$ pair production
in association with two light-quark jets, i.e., $pp \to W^{\pm}W^{\pm} q q^{\prime}+ X$,
followed by the subsequent leptonic $W$ decays.
We apply the exclusive event selection criteria, defined as
\begin{eqnarray}
p_T^{j_1}> 30~{\rm GeV},~~~~~~  p_T^{j_2}> 30~{\rm GeV},
\end{eqnarray}
to select the events with two hard jets.
In order to investigate the possibility of discriminating the signal of
the same-sign $T$-odd mirror quark pair production from its SM background,
we define $H_T$ as the scalar sum of the transverse momenta of the two
same-sign leptons, two hard jets and $\slashed{E}_T$,
\begin{eqnarray}
H_T = |\vec{p}_T^{\ell_1}| + |\vec{p}_T^{\ell_2}| + |\vec{p}_T^{j_1}| + |\vec{p}_T^{j_2}| + |\vec{p}_T^{{\rm miss}}|,
\end{eqnarray}
for a given $\ell^{\pm} \ell^{\prime \pm} + 2~jets + \slashed{E}_T$ event,
and follow the way used in searching for the top quark at
the Tevatron \cite{HTparameter}.
This kinematic variable is expected to be helpful for selecting the signal from
the SM background.

\par
The normalized $H_T$ distributions for both the signal process
$pp \to q_-q^{\prime}_- \to \ell^{\pm} \ell^{\prime \pm} + 2~jets + \slashed{E}_T + X$
at the QCD NLO and the SM background process
$pp \to W^{\pm}W^{\pm} q q^{\prime} \to \ell^{\pm} \ell^{\prime \pm} + 2~jets + \slashed{E}_T + X$
at the LO are displayed in Fig.\ref{HT}(a). We can see from the figure
that $H_T$ distributions for the signal and SM background peak at $H_T \sim 1.35~{\rm TeV}$
and $H_T \sim 300~{\rm GeV}$, respectively, and it is possible to distinguish between
the desired signal of the same-sign $T$-odd mirror quark pair production
and the SM background by adopting proper cut on $H_T$. In Fig.\ref{HT}(b)
we plot the $K$-factor of the $H_T$ distribution for the signal process.
It shows that the NLO QCD correction to the $H_T$ distribution is significant,
particularly in the range of $H_T>1.7~{\rm TeV}$. The $K$-factor varies from $0.80$ to
$0.51$ with the increment of $H_T$ from $1~{\rm TeV}$ to $2.5~{\rm TeV}$.
By comparing Figs.\ref{pt}-\ref{Mll_ET} and Fig.\ref{HT} we can see that the QCD correction
depends strongly on the event selection scheme. Our numerical results show that the NLO QCD
correction is positive in the inclusive scheme due to large positive contributions
from the real gluon/light-quark emission processes, while it suppresses the LO $H_T$ distribution
in the exclusive scheme.

\par
\section{Summary}
In this paper we calculate the same-sign $T$-odd mirror quark pair production
in the LHT at the $\sqrt{s}=14~{\rm TeV}$ LHC up to the QCD NLO.
The theoretical uncertainties
from the factorization/renormalization scale and PDFs are investigated.
We find that the NLO QCD corrections reduce the scale uncertainty of
the integrated cross section for the $T$-odd mirror quark pair production
of the first two generations significantly. The upper and lower relative
scale uncertainties of the total cross section at the central scale are
$\left(^{+19.0\%}_{-14.3\%}\right)$ at the LO, and are reduced to
$\left(^{+2.0\%}_{-4.5\%}\right)$ at the QCD NLO.
We present the dependence of the integrated cross section on the global symmetry
breaking scale $f$ and the $T$-odd mirror quark Yukawa coupling $\kappa$.
We also provide the LO and NLO QCD corrected transverse momentum and rapidity
distributions of the final products, including the leading lepton, second lepton,
leading jet and missing energy, and the invariant mass distributions of the final
produced same-sign lepton pair.
Comparing the $H_T$ distributions for the
$pp \to q_-q^{\prime}_- \to \ell^{\pm} \ell^{\prime \pm} + 2~jets + \slashed{E}_T + X$
and $pp \to W^{\pm}W^{\pm} q q^{\prime} \to \ell^{\pm} \ell^{\prime \pm} + 2~jets + \slashed{E}_T + X$
processes, we find that the signal of the same-sign $T$-odd mirror quark pair
production can be discriminated from the SM background via
$\ell^{\pm} \ell^{\prime \pm} + 2~jets + \slashed{E}_T$ final state by adopting
proper cut on the $H_T$ parameter.

\vskip 5mm
\par
\noindent{\large\bf Acknowledgments:} This work was supported
by the National Natural Science Foundation of China (Grants.
No.11275190, No.11375008, No.11375171).


\vskip 5mm
\begin{thebibliography}{99}
\bibitem{s1} 
  S. L. Glashow, Nucl. Phys. {\bf 22}, 579 (1961);
  S. Weinberg, Phys. Rev. Lett. {\bf 19}, 1264 (1967);
  A. Salam, Proc. 8th Nobel Symposium Stockholm 1968, ed. N. Svartholm
  (Almquist and Wiksells, Stockholm 1968) p.367;
  H. D. Politzer, Phys. Rept. {\bf 14}, 129 (1974).

\bibitem{s2} 
  P. W. Higgs, Phys. Lett. {\bf 12}, 132 (1964), Phys. Rev. Lett. {\bf 13}, 508 (1964), Phys. Rev. {\bf 145}, 1156 (1966);
  F. Englert and R. Brout, Phys. Rev. Lett. {\bf 13}, 321 (1964);
  G. S. Guralnik, C. R. Hagen and T. W. B. Kibble, Phys. Rev. Lett. {\bf 13}, 585 (1964);
  T. W. B. Kibble, Phys. Rev. {\bf 155}, 1554 (1967).

\bibitem{fine-tuning} 
  R. Barbieri and A. Strumia, Phys. Lett. {\bf B462}, 144 (1999).

\bibitem{HiggsA} 
  G. Aad {\it et al.} (ATLAS Collaboration), Phys. Lett. {\bf B716}, 1  (2012).

\bibitem{HiggsC} 
  S. Chatrchyan {\it et al.} (CMS Collaborations), Phys. Lett. {\bf B716}, 30 (2012).

\bibitem{Little Higgs models} 
  N. Arkani-Hamed, A. G. Cohen, and H. Georgi, Phys. Lett. {\bf B513}, 232 (2001);
  M. Schmaltz and D. Tucker-Smith, Annu. Rev. Nucl. Part. Sci. {\bf 55}, 229 (2005);
  M. Perelstein, Prog. Part. Nucl. Phys. {\bf 58}, 247 (2007), and references therein.

\bibitem{the LH} 
  N. Akarni-Hamed, A. G Cohen, E. Katz, and A. E. Nelson, JHEP {\bf 07} (2002) 034.

\bibitem{EWconstraints} 
  C. Csaki, J. Hubisz, G. D. Kribs, P. Meade and J. Terning, Phys. Rev. {\bf D67}, 115002 (2003), Phys. Rev. {\bf D68}, 035009 (2003);
  J. L. Hewett, F. J. Petriello and T. G. Rizzo, JHEP {\bf 10} (2003) 062;
  M. C. Chen and S. Dawson, Phy. Rev. {\bf D70}, 015003 (2004);
  W. Kilian and J. Reuter, Phys. Rev. {\bf D70}, 015004 (2004);
  Z. Han and W. Skiba, Phys. Rev. {\bf D71}, 075009 (2005).

\bibitem{$T$-parity1} 
  H. -C. Cheng and I. Low, JHEP {\bf 09} (2003) 051.

\bibitem{$T$-parity2} 
  H. -C. Cheng and I. Low, JHEP {\bf 08} (2004) 061.

\bibitem{$T$-parity3} 
  I. Low, JHEP {\bf 10} (2004) 067.

\bibitem{LHTindetail} 
  J. Hubisz and P. Meade, Phys. Rev. {\bf D71}, 035016 (2005);
  J. Hubisz, P. Meade, A. Noble and M. Perelstein, JHEP {\bf 01} (2006) 135.

\bibitem{LHTlimits} 
  J. Reuter, M. Tonini and Maikel de Vries,  JHEP {\bf 02} (2014) 053.

\bibitem{LHT-ph1} 
  A. Belyaev, C. -R. Chen, K. Tobe and C. -P. Yuan, Phys. Rev. {\bf D74}, 115020 (2006).

\bibitem{LHT-ph2} 
  D. Choudhury, D. K. Ghosh and S. K. Rai, JHEP {\bf 07} (2012) 013;
  D. Choudhury and D. K. Ghosh, JHEP {\bf 08} (2007) 084;
  S. Mukhopadhyay, B. Mukhopadhyaya and A. Nyffeler, JHEP {\bf 05} (2010) 001.

\bibitem{DuSM} 
  S. -M. Du, L. Guo, W. Liu, W. -G. Ma and R. -Y. Zhang, Phys. Rev. {\bf D86}, 054027 (2012).

\bibitem{LHT-ph3-L.W} 
  W. Liu, R. -Y. Zhang, L. Guo, W. -G. Ma and L. -W. Chen, Phys. Rev. {\bf D87}, 034034 (2013).

\bibitem{LHT-ph4} 
  R. -Y. Zhang, H. Yan, W. -G. Ma, S. -M. Wang, L. Guo and L. Han, Phys. Rev. {\bf D85}, 015017 (2012);
  X. -D. Yang, S. -J. Xiong, W. -G. Ma, R. -Y. Zhang, L. Guo and X. -Z. Li Phys. Rev. {\bf D89}, 014008 (2014).

\bibitem{CMS ssdl} 
  CMS Collaboration, JHEP {\bf 06} (2011) 077, JHEP {\bf 01} (2014) 163,
  Phys. Rev. Lett. {\bf 109}, 071803 (2012).

\bibitem{ATLAS ssdl} 
  ATLAS Collaboration,  Phys. Rev. Lett. {\bf 108}, 241802 (2012).

\bibitem{Feynman rules} 
  M. Blanke, A. J. Buras, A. Poschenrieder, S. Rechsiegel, C. Tarantino, S. Uhlig and A. Weiler,
  JHEP {\bf 01} (2007) 066.

\bibitem{FeynArts} 
  T. Hahn, Comput. Phys. Commun. {\bf 140}, 418 (2001).

\bibitem{FormCalc} 
  T. Hahn, M. Perez-Victoria, Comput. Phys. Commun. {\bf 118}, 153 (1999).

\bibitem{ellis}
 R. K. Ellis and G. Zanderighi, JHEP {\bf 02} (2008) 002.

\bibitem{hooft}
 G. t'Hooft and M. Veltman, Nucl. Phys. {\bf B153}, 365 (1979).

\bibitem{denner1991}
 A. Denner, U. Nierste and R. Scharf, Nucl. Phys. {\bf B367}, 637 (1991).

\bibitem{denner2003}
 A. Denner and S. Dittmaier, Nucl. Phys. {\bf B658}, 175 (2003).

\bibitem{TCPSS} 
  B. W. Harris and J. F. Owens, Phys. Rev. {\bf D65}, 094032 (2002).

\bibitem{KLN} 
   T. Kinoshita, J. Math. Phys. {\bf 3}, 650 (1962);
   T. D. Lee and M. Nauenberg, Phys. Rev. {\bf 133}, B1549 (1964).

\bibitem{cteq6} 
  J. Pumplin, D. R. Stump, J. Huston, H. -L. Lai, P. Nadolsky and W. -K. Tung, JHEP {\bf 07} (2002) 012;
  P. M. Nadolsky, H.-L. Lai, Q. -H. Cao, J. Huston, J. Pumplin, D. Stump, W. -K. Tung and C. -P. Yuan, Phys. Rev. {\bf D78}, 013004 (2008).

\bibitem{PDG2012} 
  J. Beringer {\it et al.} (Particle Data Group), Phys. Rev. {\bf D86}, 010001 (2012).

\bibitem{pdf uncertainty}
 G. Watt, JHEP {\bf 09} (2011) 069.

\bibitem{jet algorithm}
 G. P. Salam, Eur. Phys. J. {\bf C67}, 637 (2010).

\bibitem{HTparameter}
 S. Abachi {\it et al.} (D0 Collaboration), Phys. Rev. Lett. {\bf 74}, 2632 (1995);
 F. Abe {\it et al.} (CDF Collaboration), Phys. Rev. Lett. {\bf 74}, 2626 (1995).

\end{thebibliography}
\end{document}